# Joint Survey Processing of LSST, Euclid and WFIRST: Enabling a broad array of astrophysics and cosmology through pixel level combinations of datasets

**Consideration Areas**: Ground-based project (Medium), Infrastructure Activity, Technological Development Activity

Corresponding Author: Ranga Ram Chary; rchary@caltech.edu, +1-626-354-3931

Endorsers:
Gabriel Brammer (U. Copenhagen)
Peter Capak (Caltech/IPAC)
William Dawson (LLNL)
Andreas Faisst (Caltech/IPAC)
Sergio Fajardo-Acosta (Caltech/IPAC)
Henry C. Ferguson (STScI)
Carl J. Grillmair (Caltech/IPAC)
George Helou (Caltech/IPAC)
Shoubaneh Hemmati (JPL)
Anton Koekemoer (STScI)
BoMee Lee (Caltech/IPAC)
Robert Lupton (Princeton)
Sangeeta Malhotra (NASA/GSFC)
Peter Melchior (Princeton)
Ivelina Momcheva (STScI)
Jeffrey Newman (U. Pittsburgh)
Joseph Masiero (JPL)
Roberta Paladini (Caltech/IPAC)
Abhishek Prakash (Caltech/IPAC)
Jason Rhodes (JPL)
Benjamin Rusholme (Caltech/IPAC)
Michael Schneider (LLNL)
Nathaniel Stickley (Caltech/IPAC)
Arfon Smith (STScI)
Michael Wood-Vasey (U. Pittsburgh)
G. Bruce Berriman (Caltech/IPAC)

**Abstract**

Joint survey processing (JSP) is the pixel level combination of LSST, Euclid, and WFIRST datasets. By combining the high spatial resolution of the space-based datasets with deep, seeing-limited, ground-based images in the optical bands, systematics like source confusion and astrometric mismatch can be addressed to derive the highest precision optical/infrared photometric catalogs. This white paper highlights the scientific motivation, computational and algorithmic needs to build joint pixel level processing capabilities, which the individual projects by themselves will not be able to support. Through this white paper, we request that the Astro2020 decadal committee recognize    the JSP effort as a multi-agency project with the natural outcome being a collaborative effort among groups which are normally supported by a single agency. JSP will allow the U.S. (and international) astronomical community to manipulate the flagship data sets and undertake innovative science investigations ranging from solar system object characterization, exoplanet detections, nearby galaxy rotation rates and dark matter properties, to epoch of reionization studies. It will also result in the ultimate constraints on cosmological parameters and the nature of dark energy, with far smaller uncertainties and a better handle on systematics than by any one survey alone.

**Scientific Justification**

The Euclid, LSST and WFIRST projects will undertake flagship optical/near-infrared surveys in the next decade (Figure 1). They will map thousands of square degrees of sky and cover the electromagnetic spectrum between 0.3 and 2.0 microns with sub-arcsec resolution. This will result in the detection of several tens of billions of sources, enabling a wide range of astrophysical investigations by the astronomical community and providing unprecedented constraints on the nature of dark energy and dark matter. High-level science goals which benefit from JSP are too numerous to detail here [JSP Final Report 2019, Chary et al. Astro2020 Science White Paper #55] but are summarized below.

*Cosmology*
1. High precision photometric redshifts for galaxy clustering studies through the generation of precise multi-wavelength catalogs.
2. Joint shear analysis for weak lensing by combining the high spatial resolution space-based datasets with the deep, ground-based optical datasets.
3. Improving the identification of Type Ia supernovae (SNe) and their classification based on the properties of the host galaxies in which they occur.
4. Strong lensing time delays for background transients (e.g. QSOs and SNe) and substructure in the dark matter distribution (e.g/ H0LiCOW project, Suyu et al. 2017).

*Reionization and Galaxy Evolution*
1. Accurate resolution-matched color estimation to enable selection of epoch of reionization galaxies and quasars, and to measure their contribution to the $z > 6$ reionization budget.
2. Tracking color gradients and morphologies of LSST-selected galaxies to measure their impact on shear measurements, and the growth of stellar mass in galaxies with cosmic time.

3. Deblending the spectra of line emitters (particularly Ly-a emitters) that will be seen in the Euclid/WFIRST grism data.

*Microlensing*
1. Parallax measurements between the second Earth-Sun Lagrange point L2 (Euclid, WFIRST) and the Earth (LSST) which allow for a better handle on distances to the lens.
2. Measuring the 3D dust extinction structure of the target fields through stellar colors for more precise measurements of distance to the lens and the source
3. Using space-resolution data to alleviate source confusion in dense bulge fields to detect microlensing events at the shot-noise limit of the ground-based images.

*Stellar and solar system object motions*
1. Provide a decade-long time baseline to measure the proper motions and parallaxes of stars/solar-system objects at depths that are ~5-6 magnitudes fainter than Gaia (G < 21 AB mag). These can be used to measure kinematics of stars in streams and substructures in the Galactic halo, and in nearby galaxies to probe the nature of dark matter therein (e.g. Koposov et al. 2019).
2. Measure the trajectories and chemical composition of Solar System objects. The optical/NIR albedo is proportional to an object's composition – combining the datasets allows the shapes of small SSOs to be reconstructed from sparsely sampled light curve photometry using light curve inversion techniques (e.g. the Database of Asteroids Models from Inversion Techniques [DAMIT]; Durech et al. 2010, A&A, 513, A46; Durech et al. 2018, A&A, 617, A57).

The JSP effort described here does not aim to reach the goals themselves, but will instead provide the common data products and tools needed to facilitate community-led investigations without duplication of effort. The JSP effort serves two high-level objectives: 1) provide precise concordance multi-wavelength images (both single epoch and coadds) and catalogs over the entire sky area where these surveys overlap, and 2) provide a science platform to analyze multi-epoch concordance images and catalogs to enable a wide range of astrophysical science goals to be formulated and addressed by the research community. Given the data volumes (~100PB) and computing cycles required (~1 billion CPU hours) to achieve optimal combination of the data sets, it is most cost effective to develop these capabilities as a coherent, robust service for all users by building on the insights and expertise of the three survey projects working closely with a dedicated JSP Core Team.

It is worth noting that previous combinations of space-quality and ground-based datasets have been undertaken (e.g. HST and Subaru), but are neither at the level of precision required across bands nor over comparable survey areas which span a wide range of Galactic extinction and stellar density environments.   Depending on whether one considers only the single band optical data or all space-resolution NIR data, the combined survey area is ~2 sq. deg (e.g. COSMOS, Scoville et al. 2007, Momcheva et al. 2017 or 0.25 sq. deg, respectively (e.g. CANDELS,

Koekemoer et al. 2011), several orders of magnitude smaller than what JSP will analyze. The key problems the analysis of the survey data has faced are:
1. Astrometric offsets of ~0.1-0.2" which have only been found post-Gaia and thereby manifest themselves in an increased scatter in photometry;
2. Source confusion (Figure 2 and 3) in seeing-limited data, which has resulted in the development of prior-based source deblending and fitting tools such as APHOT, TPHOT and TRACTOR (e.g. Merlin et al. 2016 and 2019, Lang et al. 2016);
3. Quasi-arbitrary zero point offsets applied to photometry in individual bands to force better agreements in photometric redshift (e.g. Brammer et al. 2008, Dahlen et al., 2013). This is likely because of inconsistent photometry and/or ignorance of galaxy spectral energy distributions used for deriving redshifts but is extremely relevant for weak-lensing studies, which have stringent bias requirements on the redshift (e.g. Drlica-Wagner et al. 2018)

Addressing these issues to yield the ultimate cosmological, astrophysical and time-domain science will require "joint survey processing" (JSP) functionality at the pixel level that is outside the scope of the individual survey projects.

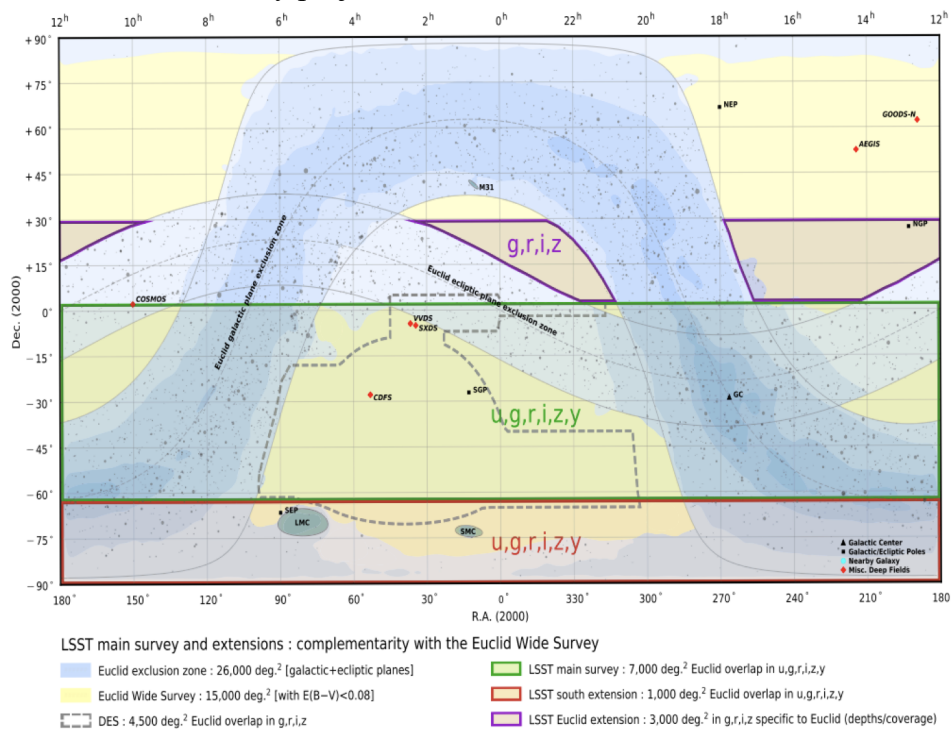

Figure 1: The sky coverage of the Euclid wide-area survey (yellow) and the possible LSST northern extension (purple rectangle, from Rhodes et al. 2017). LSST, will in addition, cover the entire 20000 deg$^2$ Southern hemisphere to ~27 AB mag over 10 years (green and red rectangles). The WFIRST high latitude survey field (2200 deg$^2$) is a part of the yellow region between RA of 300 and 90 degrees in the Southern sky while the WFIRST microlensing survey will be in the vicinity of the Galactic Center (labeled GC at 270,-30 deg). Figure courtesy of Jean-Charles Cuillandre/Euclid project. The large overlap in sky coverage between these three projects and the desire for matched precision photometry, motivates a joint analysis of the data.

## Approach

Addressing the science goals outlined above requires a data processing system that ingests images and spectra from multiple surveys (Tier 0); consolidates their astrometric and photometric definitions in a concordance frame (Tier 1); uses color and location-dependent PSFs and morphologies of sources in the concordance images to generate cross-project concordant photometric and shape catalogs (Tier 2); and exposes the concordance data and JSP catalogs to the astronomical community for astronomical manipulation and follow-up studies (Tier 3). This is summarized in Table 1.

We should emphasize that none of the tasks we outline below are intended to duplicate work already performed by the individual survey projects (e.g. instrumental calibrations, or dedicated analysis tools). The goal here is to leverage this work to construct a high-fidelity concordance data set and provide online application program interfaces (APIs) for the astronomical community to carry out novel analyses on the concordance dataset. Furthermore, while we hope for a formal data sharing agreement between the surveys, all efforts described here can be executed with public data sets that are going to be released by each of the survey projects, as per their data release schedule (Table 2).

## Overview of JSP Software Elements

The software development for JSP can be broken down into two classes: systems-level infrastructural software development and data processing software development.

The infrastructural software development involves 1) setting up and maintaining version-controlled software containers, 2) managing network throughput to ensure efficient transfer of data from storage sites to high performance computing nodes, 3) extracting unified metadata from the individual datasets, 4) providing seamless, over-the-network, data access from multiple archives through a data broker, 5) the augmentation of data visualization tools such as the LSST Science User Suite for handling multi-project datasets, and 6) the use of security and privacy software to ensure that i) a user does not overwhelm computational resources, ii) a malicious user does not alter the contents of the archives, and iii) the work of an individual user is private and cannot be viewed by others.

The data processing development work involves 1) the creation of software to perform the alignment and standardization of the reduced data products, 2) generation of the concordance optical/NIR images and catalogs including color, time, brightness and location-dependent point spread functions, 3) development of software that enables end-users to perform custom manipulation of the concordance datasets (e.g., creating custom coadds, non-sidereal stacking, and point source fitting), 4) the generation of high spatial resolution extinction maps and a color correction tool, 5) performing cross-calibration and astrometry checks, and 6) creating a means of injecting fake sources of different shapes after convolving with the corresponding point spread function to assess basic quantities like completeness, reliability and photometric biases as a function of morphology.

By containerizing the analysis and processing software, we can ensure consistency and repeatability. The exact same versions of the software will be used for creating all of the concordance products and the same container images will be made available to users of the science platform. Containerization allows us to preserve the versions of not only the software applications themselves, but also all of underlying libraries.

The data broker and client library that we will design will significantly lower the barrier to entry for algorithm developers wishing to co-analyze multi-mission, multi-wavelength data. This data broker would allow users to query by position (or survey object id) on the sky, specify a wavelength range and spatial extent, and have returned to them images (single visit or co-added) aligned to a common grid (e.g. Gaia) plus metadata drawn from the archives of appropriate surveys (e.g. LSST, WFIRST, PanSTARRS, ZTF, SDSS and GALEX). The metadata supplied with these images would include precise astrometric mapping and a model of the PSF interpretable by JSP algorithms.

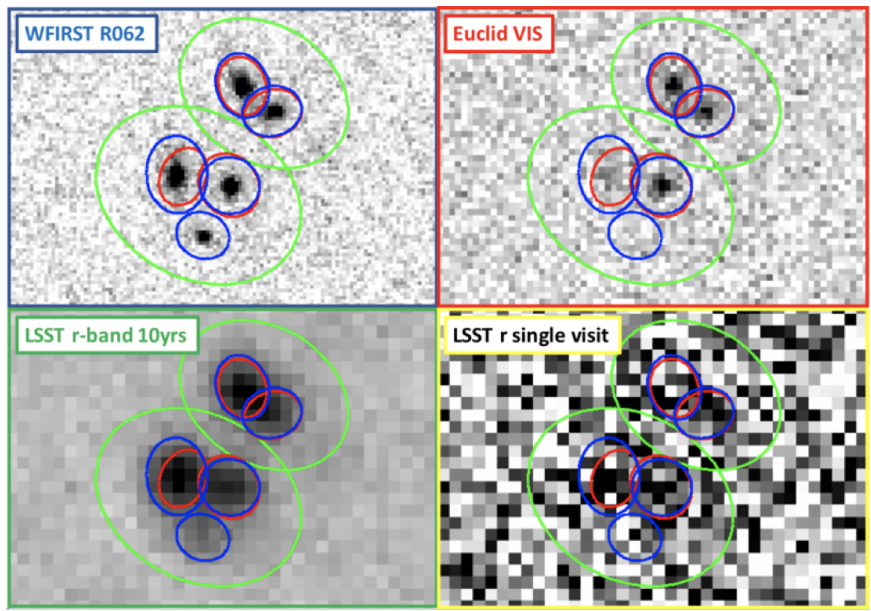

Figure 2: An illustration of source confusion in an optical band (centered at ~6000A) from the three primary surveys, along with the isophotes derived from photometry on each of the images. The green isophotes are derived from the LSST r-band full-depth data of 27.5 AB mag, the red isophotes are from the Euclid only VIS data while the blue isophotes are for the deeper WFIRST data (Table 1). The sources are barely detected in the LSST single epoch data. In the absence of the deeper, space-resolution data, source confusion would result in both erroneous shape and photometry estimates in LSST data and also affect catalog matching. Conversely, both Euclid and WFIRST rely on deconfused optical photometry from LSST to get reliable photometric redshifts for galaxies that are detected in their respective surveys.

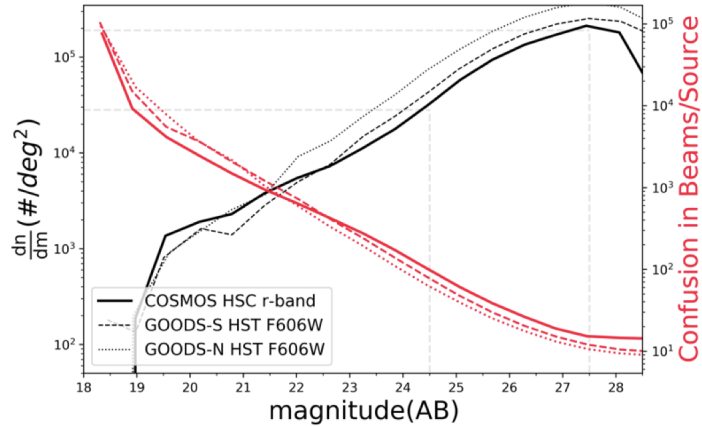

Figure 3: Visible band extragalactic source counts in the COSMOS, GOODS-S and GOODS-N fields shown as the black lines with the corresponding beams/source, a parameter of confusion, show as the red lines and corresponding to the right axis. Beams/source=40 (corresponding to ~25 AB mag) is the classical confusion limit. Using prior information on position and morphologies of sources allows one to reliably extract sources down to ~8 beams/source depending on the accuracy of the PSF and the quality of the priors (Magnelli et al. 2009, 2011). Since LSST has a single epoch photometric sensitivity of 24.5 AB mag, with ~200 visits per band over 10 years, the images will start to be affected by confusion, within a year of LSST operations, if the spatial resolution has a median seeing of 0.8-0.9" FWHM. This implies that between 10-25% of galaxies will have their photometry affected by confusion in 1 year stacks of LSST data with up to 50% affected in the full depth stacks.

**Computing Requirements**

Three types of computing runs have been identified. The first is required to produce the concordance products and package them for use by the community. The second type encompasses bulk runs, managed by large science collaborations aimed at major science investigations such as Weak Lensing or proper motion searches. The third type accounts for the more "localized" science investigations conducted by the community.

The generation of concordance products will need to be run roughly annually to remain up-to-date with releases from the projects. They would start in early 2023, as soon as overlapping data from both LSST and Euclid are publicly available. We estimate that this effort would involve a total of about 330 million CPU hours, combining both products and supporting simulations. However, each iteration of this for the intermediate data releases is likely only about a tenth of that, since the overlapping area will be small. It is reasonable to assume that this could be conducted at the DOE supercomputing centers as a contribution to the JSP effort. In addition, generating the single-epoch simulations, PSFs and catalogs is estimated to account for ~600 million CPU hours.

Bulk computing runs for major investigations are also likely to be needed roughly annually starting in late 2023 as surveys cover more sky or go deeper. We assume these would be run at supercomputing centers or potentially on commercial clouds if the costs are reasonable. These runs would have to be scheduled and staged, with test runs to validate the resourcing leading up to production runs. They would be coordinated with the projects, since they will require streaming major amounts of data out of the project archives, and may impact on-going science operations or frustrate access by other users. The resources required for these computing runs, estimated at tens of millions of CPU hours, will be procured by the investigation teams and are not budgeted here.

The science investigations conducted on a more modest scale by community research groups will run wherever those investigators can accommodate their tailored computing runs built on JSP software containers. We do not propose a centralized facility to host this computing, and assume the individual investigations will be funded to cover computing costs.

Table 1: Tiers of Data and Software Products from JSP

|        | | | | |
|--------|---|---|---|---|
| Tier 0 | Single epoch frames and metadata [CoR1, GSR1, SSO1] | Single-frame astrometric solution [CoR1, SSO2] | Color/location dependent PSFs [CoR3, GER2] | Photometric calibration parameters [CoR1] |
| Tier 1 | Concordance solutions for astrometry and photometry [CoR2, GSR1] | Optimal coadds, noise and coverage maps [CoR4, GER2] | Interface to external datasets (e.g. Gaia) and PSFs for forced photometry [CoR6] | Astrometrically aligned spectral products [GER1] |
| Tier 2 | Joint photometry tools, catalogs, residual maps [CoR3] | High resolution Galactic extinction maps [CoR7, MLR2, GSR3] | Point source fitting tool [MLR1, GSR2] | |
| Tier 3 | Science platform [GER4] | Custom image coaddition tool [CoR5, SSO1] | Cross-project fake source injection software [CoR4] | Spectral extraction and decontamination tool [GER3] |

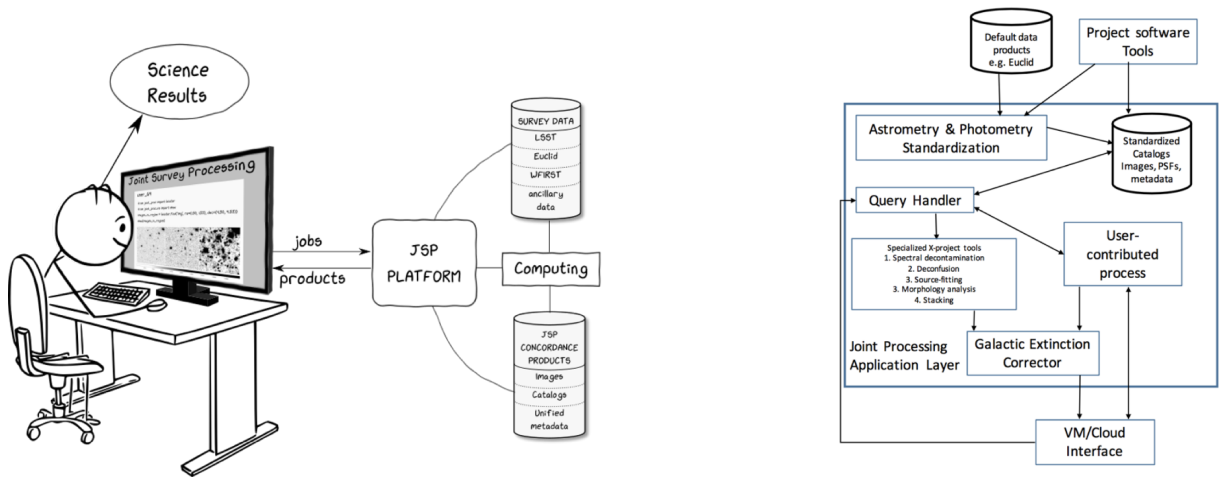

Figure 4: Strawman architecture of the JSP system. The overall layout is shown on the left while the JSP application platform layer is shown on the right. The processing over the sky area that overlaps the surveys will be executed in intervals of times (matched to project data releases) and the results will be stored in a database for easy retrieval. Customized processing applications will be executed on-the-fly by code on standardized, calibrated frames from the projects which will be stored at their corresponding data processing centers.

## Schedule

Table 2, Tasks and Milestones, shows a draft schedule, summarizing the prime mission span for each of the three projects, then listing each JSP task and its period of performance. The anticipated release dates for data from each of the three projects are shown in those lines, with DR for Data Release and QR for Quick-look Release. Active periods of JSP tasks are color-coded for development, maintenance and operations, and their nominal delivery dates and milestones are indicated by diamonds, and are tied to data releases by the three projects. This draft schedule recognizes the focused efforts required leading up to data releases after the launch of Euclid and start of full survey operations of LSST in 2022, and similarly tied to the launch of WFIRST in 2025. FY32 as a final year in this estimate is somewhat arbitrary but should nominally be funded until the final WFIRST and LSST data releases.

The tasks are divided into three categories, with Infrastructure, Concordance Images and Catalogs and Ancillary Science Tools, each having the corresponding rows color-coded.

## Cost Estimates and Agency Funding

There are two main objectives for JSP, the first being the creation of concordance images and catalogs in the overlapping areas of sky and the second being the creation of a science platform to allow users to perform customized analysis, using the concordance products. These goals are reached by implementing tasks in categories Infrastructure and Concordance Images/Catalogs, so it is reasonable to allocate those costs to the JSP Core Team and survey projects. The Ancillary Science Tools category is aimed at supporting or developing additional tools for research, and is therefore allocated primarily to the community, with some support by JSP Core Team.

Summing over the years FY19 to FY24, we estimate a total effort of ~100 WY. For FY25-FY32, the total effort is ~104 WY. Weighting by the effort distribution in each task, and summing over all years, we estimate a total of 204 WY, 69% of which is allocated to JSP Core Team, 15% to the individual survey projects, and 16% to the community. The 204 WY number is largely driven by the duration of over a decade. **During the five peak years, the effort averages about 24 WY/year for all categories and sources of funding.**

Given its nature as a cross-project augmentation to the science, JSP has been envisaged from the start as funded by a combination of DOE, NASA and NSF, as additional scope to their funding the construction of the three projects. The three agencies have been aware of JSP formulation activities, and discussing them through their "Tri-Agency Group." **This White Paper therefore proposes that JSP be recognized explicitly as a multi-agency project with the natural result being a collaborative effort among communities usually supported exclusively by one or the other agency.** Given the range of cost per WY at different institutions, the total JSP labor effort from start to FY32 is much larger than a typical science grant but a third of the size of the smallest Explorer mission.

With the above in mind, we propose a notional distribution of funding among the agencies, that is inspired by the contents of the various tasks. A reasonable scenario is that community effort would be funded by proposal and peer-review rounds, with NSF carrying a major portion. The individual survey project participation would be funded by their respective funding agencies based on requests by each project. The JSP Core Team, with membership drawn from multiple institutions, could be constituted through direct funding by NASA to one or more of its centers, with contributions from DOE and NSF in the form of dedicated funds or of personnel direction.

A major cost component for JSP is computing, as detailed above. We estimate a billion or more CPU hours will be needed. A reasonable scenario would be for DOE and NSF to allocate that computing load on their Supercomputing Centers, or procure the equivalent on the commercial cloud if that became a competitively priced, viable option in the next few years. This would motivate more engagement in JSP by DOE laboratory staff, thus helping the overall effort.

## References

- Chary, R., et al., 2019, Astro2020 Science White Paper #55, **https://tinyurl.com/y2h7ch6s**
- Dahlen, T., et al., 2013, ApJ, 775, 93
- Drlica-Wagner, A., et al., 2018, ApJ, 235, 33 (DES Collaboration)
- Durech, J., et al. 2010, A&A, 513, 46
- Durech, J., et al., 2018, A&A, 620, 91
- Koekemoer, A. M. et al. 2011, ApJS, 197, 36
- Koposov, S., et al., 2019, MNRAS, 485, 4726
- Lang, D., et al., 2016, Astrophysics Source Code Library, record ascl:1604.008
- Lee, K.-S., et al., 2012, ApJ, 752, 66
- Magnelli, B., et al., 2009, A&A, 496, 57
- Magnelli, B., et al., 2011, A&A, 528, 35
- Merlin, E., et al., 2016, A&A, 595, 97
- Merlin, E., et al., 2019, A&A, 622, 169
- Momcheva, I., et al., 2017, PASP, 129, 5004
- Rhodes, J., et al., 2017, ApJS, 233, 21
- Scoville, N., et al., 2007, ApJS, 172, 38
- Suyu, S., et al., 2017, MNRAS, 468, 2590

Table 2: Task Development Schedule with Delivery Milestones